# Quantum sensing via magnetic-noise-protected states in an electronic spin dyad

*Carlos A. Meriles[†], Pablo R. Zangara, and Daniela Pagliero*

Extending the coherence lifetime of a qubit is central to the implementation and deployment of quantum technologies, particularly in the solid-state where various noise sources intrinsic to the material host play a limiting role. Here, we theoretically investigate the coherent spin dynamics of a hetero-spin system formed by a spin $S = 1$ featuring a non-zero crystal field and in proximity to a paramagnetic center $S' = 1/2$. We capitalize on the singular energy level structure of the dyad to identify pairs of levels associated to magnetic-field-insensitive transition frequencies, and theoretically show that the zero-quantum coherences we create between them can be remarkably long-lived. Further, we find these coherences are selectively sensitive to 'local' — as opposed to 'global' — field fluctuations, suggesting these spin dyads could be exploited as nanoscale gradiometers for precision magnetometry or as probes for magnetic-noise-free electrometry and thermal sensing.

## 1. Introduction

Paramagnetic color centers in wide bandgap semiconductors are attracting broad interest as a platform for quantum information processing in the solid state, most notably, due to their favorable spin properties[1]. Indeed, the relative robustness of spin angular momentum as compared to other degrees of freedom often translates into long coherence lifetimes[2,3], even at room temperature[4]. Magnetic fluctuations from the environment — e.g., created by the surrounding bath of electronic and nuclear spins — often set a limit on the time duration of these coherences, the reason why much effort has been devoted to reduce their impact. Adding to "static" strategies — relying on higher sample purity and selective depletion of nuclear-spin-active isotopes from the host crystal — "dynamical" schemes have been developed that effectively decouple the spin qubit from its environment, hence leading to extended coherence times[5]. These methods are proving useful not only in the context of quantum information processing but also for metrology, where they are being exploited to selectively highlight interactions otherwise obscured through quick relaxation of the probe.

One complementary strategy to extending the qubit coherent evolution is to tune the Hamiltonian to render the dynamics selectively insensitive to deleterious sources of noise, most importantly, magnetic fluctuations[6]. This is usually attained by bringing the system to conditions where the energies $E_1$, $E_2$ of two eigenstates become unresponsive to changes in the magnetic field (i.e., where $\partial E_{1,2}/\partial B \sim 0$). While fluctuation-insensitive spin dynamics have been observed at zero or low magnetic fields[7-11], level anti-crossings at higher fields often provide an alternative route to mitigating the effects of magnetic noise; the degree of protection depends on the curvature of the energy levels at the anti-crossing and hence disappears for sufficiently strong detuning. This "parametric" approach has been extensively applied to atomic clocks where hyperfine transitions are used as standards[12]. More recently, similar ideas have been adapted to solid state systems, including Bi and P donors in silicon[13,14], N impurities in diamond[7], and rare-earth dopants in garnets[15,16]; tuned Hamiltonians have also been developed to protect superconducting qubits from charge, flux, or current noise[17,18].

While the methods above typically build on the properties of individual atom-like systems, optimal control techniques[19] and, most notably, entanglement[20] between individually addressable qubits provide yet another, arguably less explored path to enhanced sensing. For example, recent work with a spin-active color center hyperfine-coupled to neighboring nuclear spins capitalized on entangled states to demonstrate magnetic field

C.A. Meriles, D. Pagliero
Department. of Physics,
CUNY-City College of New York,
New York, NY 10031, USA.
E-mail: cmeriles@ccny.cuny.edu

C.A. Meriles
CUNY-Graduate Center,
New York, NY 10016, USA

Pablo Zangara
Universidad Nacional de Córdoba,
Facultad de Matemática, Astronomía, Física y Computación,
Córdoba, Argentina.

Pablo Zangara
CONICET,
Instituto de Física Enrique Gaviola (IFEG), Córdoba, Argentina.

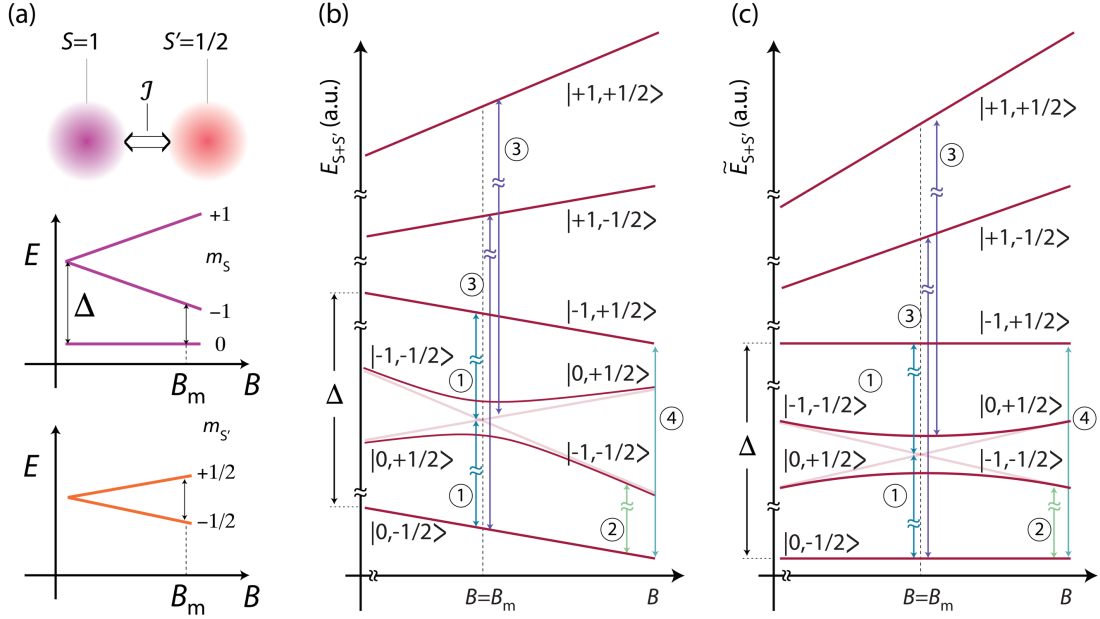

**FIG. 1: Energetics of the two-spin system.** (a) We consider a spin $S = 1$ featuring a crystal field $\Delta$ and coupled to a paramagnetic impurity $S' = 1/2$ via a dipolar interaction of amplitude $\mathcal{J}$; for all values, we assume the magnetic field direction coincides with that defined by the crystal field at spin $S$. At the level-crossing field $B_m = \Delta/2$, the energy separation between the $m_S = 0$ and $m_S = -1$ states of spin $S$ coincides with the Zeeman splitting $m_{S'} = \pm 1/2$ of $S'$ (upper and lower energy diagrams, respectively). (b) Energy diagram of the combined system. Circled numbers denote individual transitions of distinct frequencies. (c) Same as in (b) but after adding a term $\delta E = +|\gamma_e|B$ to all energy levels.

detection with precision beyond the standard quantum limit[21].

Here we theoretically study a pair of interacting electronic spins featuring different spin numbers; for concreteness, we focus on a system comprising a nitrogen-vacancy (NV) center in diamond and a proximal spin-1/2 paramagnetic impurity (such as a substitutional neutral nitrogen impurity, the so-called P1 center[22]) but we later show the ideas can be broadly generalized. We first investigate the dynamics of the two-spin system near an energy anti-crossing to show that although level bending offers first-order protection against decoherence, the mechanism is impractically vulnerable to detuning of the operating magnetic field. We capitalize on these findings, however, to subsequently create two-spin, zero-quantum coherences, and show these states are robust against global magnetic field fluctuations to all orders, regardless the operating external magnetic field. Finally, we build on the underlying physical differences between the constituent spins in the dyad to show our approach can be exploited to implement magnetic-noise-insensitive electrometry or thermometry protocols.

## 2. Protecting Quantum Coherences

### 2.1 Physical system

Fig. 1a lays out the system under consideration: Spin $S = 1$ features a state triplet with a crystal field splitting $\Delta$ while spin $S' = 1/2$ represents a dipolarly coupled paramagnetic center in its proximity; we also assume a magnetic field $B$ aligned along the quantization axis defined by the symmetry axis of spin $S$. For a field $B_m \approx \Delta/2$, the energy difference between the $m_S = 0$ and $m_S = -1$ states matches the Zeeman splitting between the $m_{S'} = \pm 1/2$ states of spin $S'$, a condition already exploited, e.g., to spin polarize the paramagnetic center and adjacent nuclei in the crystal host[23-26]. Fig. 1b shows the energy $E_{S+S'}$ eigenvalues for the combined spin system as a function of the applied magnetic field: At $B_m$, the dipolar coupling between the NV and the paramagnetic center produces a level anti-crossing between the $|0, +1/2\rangle$ and $|-1, -1/2\rangle$ branches in the diagram, whose gap is proportional to the inter-spin coupling $\mathcal{J}$. At the level anti-crossing, the two states hybridize into $|\pm\rangle = 1/\sqrt{2}\,(|0, +1/2\rangle \pm |-1, -1/2\rangle)$ and the transition frequency between the corresponding energies reaches a minimum, hence making coherences between these latter two states robust (to first order) against magnetic field fluctuations.

More interestingly, though, the energy separation between $|0, -1/2\rangle$ and $|-1, +1/2\rangle$ is independent of the applied magnetic field meaning that single-quantum coherences between either level and $|+\rangle$ or $|-\rangle$ should also be protected (transition 1 in Fig. 1b). This is better seen in Fig. 1c, where we plot the modified eigen-energies $\tilde{E}_{S+S'} = E_{S+S'} + \delta E$ for $\delta E = +|\gamma_e|B$ ($\gamma_e$ denotes the electronic gyromagnetic ratio). Near $B_m$, none of the four lower branches depends on the magnetic field, with the consequence that state superpositions near the level anti-crossing must be longer-lived. The same also applies away from $B_m$ for coherences between $|0, -1/2\rangle$ and $|-1, +1/2\rangle$, though creating superpositions between these levels is more involved as they cannot be attained via microwave (MW) excitation alone. Below we investigate the dynamics of spin coherences both near and far away from the level anti-crossing and show only the latter case provides practical levels of protection against magnetic noise. In what follows, we ignore the hyperfine coupling of either point defect with their nuclear host, a simplification justified herein given the comparatively slow nuclear spin dynamics (see Supplementary Information, Section I).

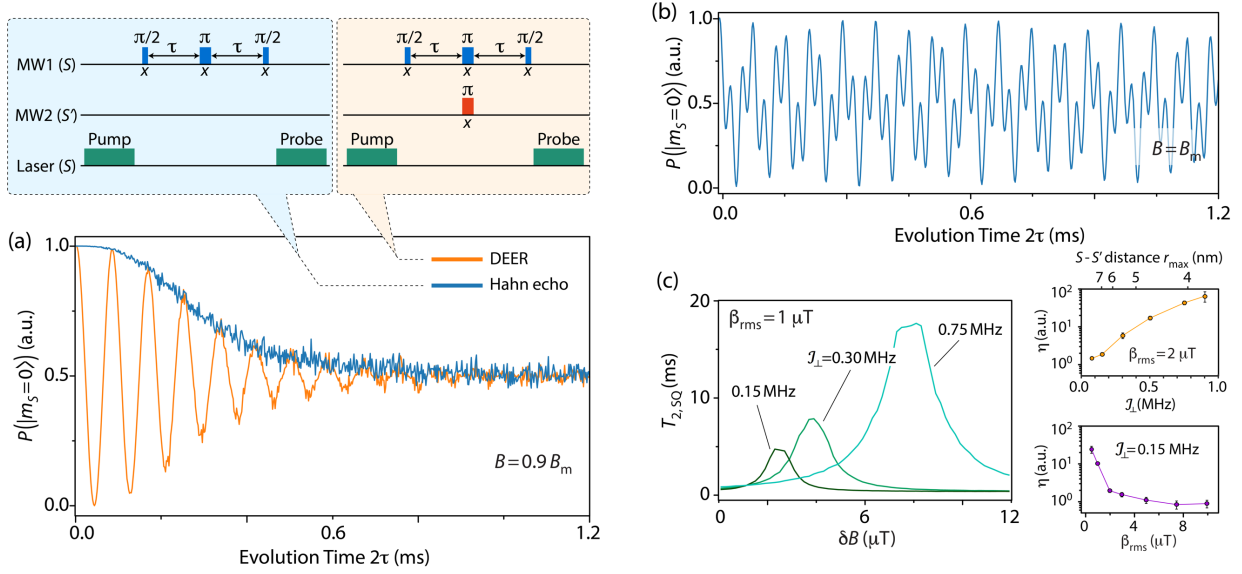

**FIG. 2: Coherence protection near the level anti-crossing.** (a) Hahn-echo and DEER coherent response of spin $S$ away from $B_m$ (b) When $B = B_m$, both protocols coincide; the plot displays the calculated response of spin $S$. In (a) and (b), $\beta_{rms} = 1$ µT and $\mathcal{J}_\perp = 0.15$ MHz. (c) Hahn-echo spin transverse relaxation time $T_{2,SQ}$ of spin $S$ as a function of the field detuning $\delta B \equiv B - B_m$ for different $S - S'$ coupling constants. The side panels display the calculated coherence lifetime enhancement $\eta$ (relative to $T_{2,SQ}$ at a field sufficiently far away from $B_m$) as a function of the inter-spin coupling and spin-noise amplitude (upper and lower plots, respectively). For reference, we express the upper horizontal axis in the upper right-hand insert in terms of the distance $r_{max}$ representing the maximum separation for color centers featuring a coupling of magnitude $\mathcal{J}$. In all cases, we assume $\mathcal{J} = \mathcal{J}_\parallel = \mathcal{J}_\perp$.

**2.2 Coherent spin dynamics of the dyad near $B_m$**

To better assess the system response to magnetic fluctuations, we start by considering a semi-classical model where magnetic noise of amplitude $\beta(t)$ stems from outside sources changing randomly over time with some characteristic rate (see Supplementary Information, Section I). Away from the level anti-crossing, the linear relation between the applied field and the transition frequencies of spins $S$, $S'$ establishes a proportionality between the decoherence rate and the fluctuator-induced root-mean-square (rms) magnetic field $\beta_{rms}$. We can therefore gauge the effectiveness of the anti-crossing as a shield against noise by comparing the system response as we approach $B_m$.

For future comparison, we first tune the MW frequency to the $|0\rangle \leftrightarrow |-1\rangle$ transition and calculate the system evolution under a Hahn-echo protocol away from the level anti-crossing (Fig. 2a); consistent with experimental practice, we assume optical initialization and readout of spin $S$ (here implicitly associated to an NV[27,28]). We then make the magnetic field equal to $B_m$ and derive the system response in the typical limit where the excitation bandwidth of all MW pulses is greater than the anti-crossing gap (Fig. 2b). Besides introducing a fast signal beating — mainly arising from a double-quantum coupling term $H_{DQ} = 2\pi \mathcal{J}_\perp (S_+ S'_+ + S_- S'_-)$ exclusively active near $B_m$, see Supplementary Information, Section I— proximity to the level anti-crossing leads to significantly longer-lived coherences (here captured through the characteristic time $T_{2,m}$ in a stretched exponential fit). Specifically, for the conditions assumed in the figure we find that the ratio $\eta \equiv T_{2,m}/T_{2,SQ}$ between the single-quantum (SQ) coherence time of spin $S$ at and away from $B_m$ can be large (depending on the dipolar coupling and noise amplitude). This ratio remains unchanged if, rather than the Hahn-echo, we take the double electron-electron resonance (DEER) protocol as the reference because both lead to the same coherence lifetimes under the simplified conditions assumed herein (Fig. 2a). Importantly, we reach the same conclusion even when we consider the differing rotation angles experienced by either spin in the dyad under a common MW field at $B_m$ (a direct consequence of the hetero-spin nature of the dyad, see Supplementary Information, Section I).

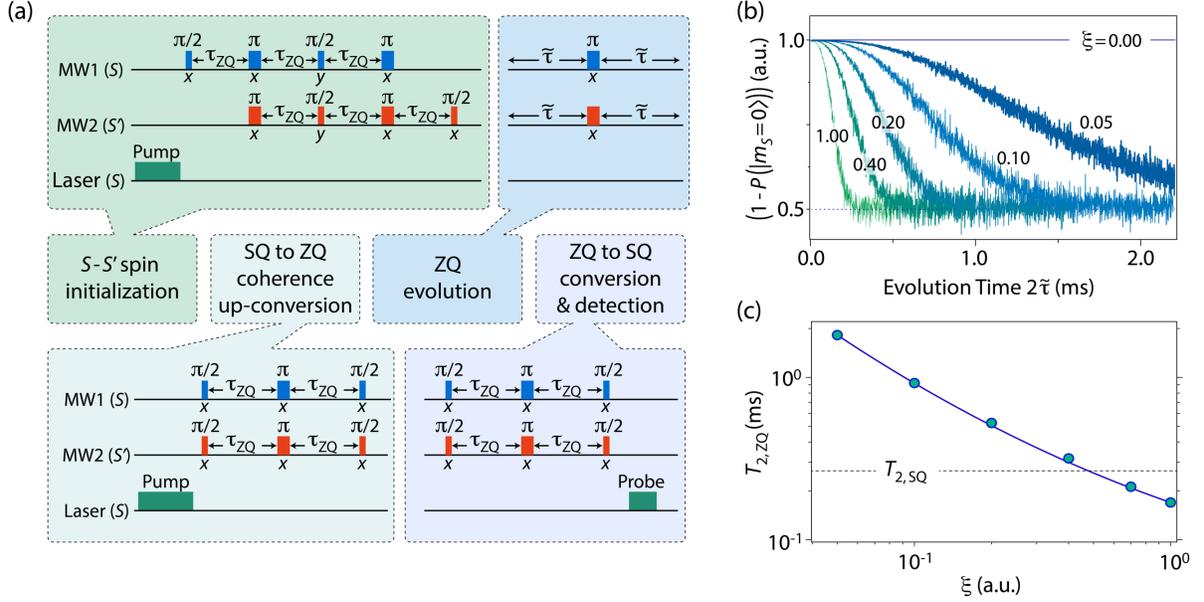

**FIG. 3: Extending coherence lifetimes by inter-spin entanglement.** (a) Spin control protocol; we choose $\tau_{ZQ} = (4\mathcal{J}_\parallel)^{-1}$. (b) Calculated system response as extracted from monitoring spin $S$ upon a zero-quantum free evolution of variable time $\tilde{\tau}$ and assuming $\beta_{rms} = 1\ \mu T$ and $\mathcal{J}_\parallel = 50$ kHz ($r_{max} \cong 10$ nm). (c) Calculated zero-quantum (ZQ) coherence time $T_{2,ZQ}$ as a function of the fractional magnetic noise difference $\xi$; the dashed line represents the single-quantum coherence lifetime of spin $S$ far from the level anti-crossing, here serving as a reference.

Numerical modeling as a function of the external field, however, indicates that noise protection is limited to a narrow band. In particular, Fig. 2c shows that even in the regime of strongly coupled dyads ($\mathcal{J}_\perp = 0.75$ MHz, corresponding to an inter-defect separation of ~4 nm), robustness against decoherence is limited to a window $\delta B_m \sim 4\ \mu T$ around $B_m$, hence making the system susceptible to slow fluctuations (e.g., stemming from temperature changes[29]).

## 2.3 Protecting spin coherences far away from $B_m$

An intriguing characteristic in the energy diagram of Fig. 1c — only exploited indirectly in Fig. 2 — is that the energy separation between the $|0, -1/2\rangle$ and $|-1, +1/2\rangle$ levels remains constant (and equal to $\Delta$) at all fields, which suggests that coherences between these levels would be intrinsically protected against global magnetic fluctuations. Since the operating magnetic field is largely inconsequential, we move $B$ away from $B_m$, and write the Hamiltonian as

$$H = \Delta S_z^2 + |\gamma_e|B(S_z + S_z') + 2\pi\mathcal{J}_\parallel S_z S_z', \quad (1)$$

where we assume the magnetic field and crystal field axis are aligned, and $\mathcal{J}_\parallel$ is the secular dipolar coupling amplitude (Supplementary Information, Section I). Note that the double-quantum term $H_{DQ}$ — responsible for the energy gap at the level crossing — becomes non-secular away from $B_m$, and can hence be ignored. Limiting our description to the manifold spanned by states $|0, +1/2\rangle$, $|-1, +1/2\rangle$, $|0, -1/2\rangle$, and $|-1, -1/2\rangle$, spin $S$ can be described through $\tilde{S}$, a fictitious spin-1/2 operator, and the Hamiltonian takes the reduced, more convenient form

$$\tilde{H} = (|\gamma_e|B - \Delta)\tilde{S}_z + (|\gamma_e|B - \pi\mathcal{J}_\parallel)S_z' + 2\pi\mathcal{J}_\parallel \tilde{S}_z S_z'. \quad (2)$$

Following the energy diagram in Fig. 1c (see transition 4), coherences between states $|-1, +1/2\rangle$ and $|0, -1/2\rangle$ must be resilient to magnetic field fluctuations. Since direct microwave excitation cannot produce this type of coherences, we use a two-step strategy where we first initialize the spin system via a multi-pulse protocol whose timing ensures full transfer of the NV polarization to spin $S'$ (see Ref. [30] and Supplementary Information, Sections II and III). After NV spin re-pumping, we use a DEER-like sequence (with inter-pulse separation $\tau_{ZQ} = (4\mathcal{J}_\parallel)^{-1}$, Fig. 3a) to transform the initial state (here expressed as an effective density matrix operator $\rho_{eff}(0) = \frac{1}{2}(\tilde{S}_z - S_z')$) into a zero-quantum coherence, namely

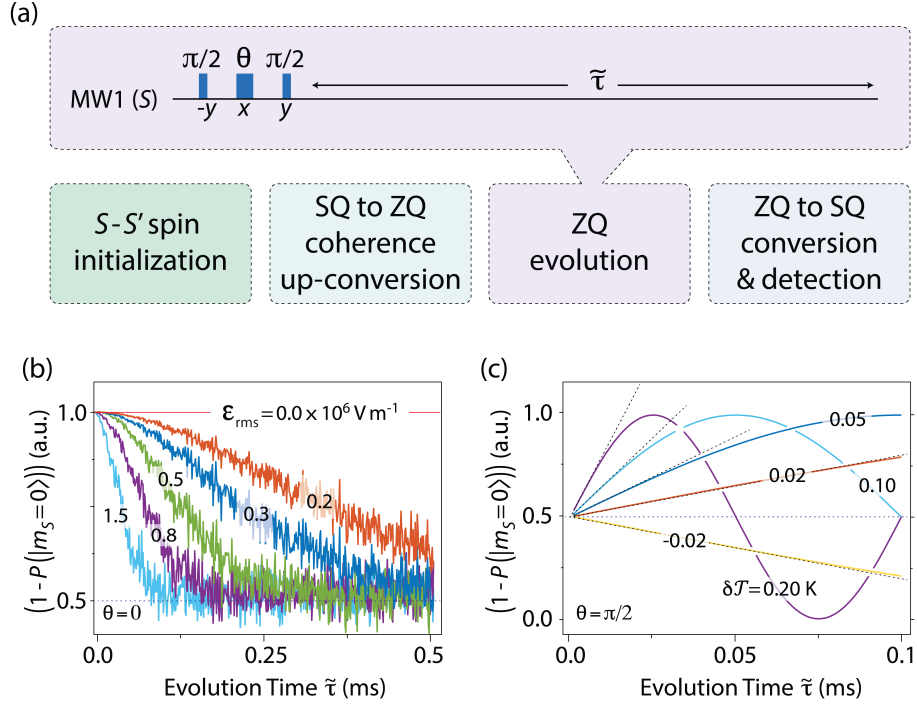

**FIG. 4: Alternative sensing modalities.** (a) Schematics of the pulse sequence; the first composite pulse amounts to a phase rotation of spin $S$ by a variable amount $\theta$. (b) Electric-field-noise-selective relaxometry; the electric field has amplitudes $\langle \mathcal{E}_z^2 \rangle = \langle \mathcal{E}_x^2 \rangle = \langle \mathcal{E}_y^2 \rangle = \mathcal{E}_{\text{rms}}^2$ and we assume a constant temperature; note that since $\theta = 0$, the sequence is identical to that in Fig. 3 except that there is no $\pi$-pulse during the zero-quantum evolution. (c) Thermal sensing modality (assuming $\mathcal{E}_{\text{rms}} = 0$); the temperature change $\delta\mathcal{T}$ can be extracted from the signal slope at early times (and the known thermal sensitivity of the NV at room temperature). In (b) and (c), $\beta_{\text{rms}} = 1\,\mu\text{T}$ and $\xi = 0$. The absence of noise in (c) reflects the suppression of the magnetic fluctuations, for simplicity, the only noise source in these calculations. All other conditions as in Fig. 3.

$$\rho_{\text{eff}}(2\tau_{\text{DQ}}) = \frac{1}{2i}\left(\tilde{S}_- S'_+ - \tilde{S}_+ S'_-\right), \quad (3)$$

where $\tilde{S}_\pm = \tilde{S}_x \pm i\tilde{S}_y$ and analogously for $S'_\pm$. Importantly, this state is insensitive to global magnetic noise of arbitrary amplitude because the phase picked-up by $\tilde{S}_\pm$ is cancelled by $S'_\mp$; we contrast this response with that observed in Fig. 2c, where the level of protection degrades as the rms noise amplitude increases. Our strategy is related to (but different from) the singlet-triplet long-lived coherences already introduced in nuclear magnetic resonance for inhomogeneity-free spectroscopy[31]. Interestingly, the magnetic-noise-resilient state in Eq. (3) is formally equivalent to the one produced via the manipulation of two spin-1/2 nuclei[32]; we show below, however, how the underlying physical differences between the two electron spins in the dyad can be exploited to enact alternative, otherwise unattainable sensing modalities.

### 3. Application to quantum metrology

Complete robustness to magnetic noise must be seen, of course, as a limit case because in practice the physical separation between the electronic spins of the dyad introduces some finite difference between the magnetic noise amplitudes $\beta_S(t)$ and $\beta_{S'}(t)$ experienced by either spin at a given time $t$. The immediate consequence is an imbalance between the instantaneous frequency shifts in each spin with the concomitant decay of the two-spin entanglement. In other words, the system behaves as a gradiometer, selectively sensitive to magnetic field fluctuations occurring on the scale of the separation between the dyad spins. Interestingly, this class of dephasing can be mitigated through the intercalation of an inversion pulse at the midpoint of the zero-quantum evolution interval, which yields an echo not unlike that characteristic in single-quantum coherences (Supplementary Information, Section IV); we include a $\pi$-pulse here to facilitate comparison with the unprotected case (Fig. 3a).

We benchmark the system response in Fig. 3b where we plot the time trace of the fractional NV population in $|m_S = 0\rangle$ — detected upon a zero- to single-quantum conversion, Fig. 3a — as a function of the zero-quantum evolution time $2\tilde{\tau}$. To better gauge the impact of the local environment, we keep the rms noise amplitude at both spin sites equal (and constant), but gradually alter the fractional contribution from local, paramagnetic-center-selective sources; we quantify these changes through the parameter

$\xi \equiv \langle(\beta_S - \beta_{S'})^2\rangle / (\langle(\beta_S)^2\rangle + \langle(\beta_{S'})^2\rangle)$, where we use brackets to indicate time average. Fig. 3c shows the extracted zero-quantum coherence lifetime $T_{2,ZQ}$ as a function of $\xi$. Noting that $\xi \to 1$ as fluctuations at either spin site become independent, the quick coherence decay we observe indicates the system behaves as a sensitive gradiometer; on the other hand, comparison with $T_{2,SQ}$ (far from $B_m$) shows that longer system lifetimes can still be attained for considerably different noise environments.

An immediate corollary to the results above is that the dyad remains selectively sensitive to environmental changes that only affect spin $S$, hence allowing one to envision alternative sensing modalities protected from global magnetic noise. Fig. 4 shows two complementary instances of quantum sensing that leverage the distinctive physical roots underlying the spin-1 nature of spin $S$. In the first example, we illustrate an application to electrometry[33,34] — here designed to expose electric noise through changes in $T_{2,ZQ}$, Figs. 4a and 4b — which we derive after including in the Hamiltonian the NV coupling terms to electric fields (ignored in Eq. (1) for simplicity, see Supplementary Information, Section IV). Since magnetic fluctuations often dominate the NV decoherence rate, the ability to remove this contribution can prove relevant to the realization of novel forms of electric-field-sensitive microscopy[35].

Fig. 4c shows an alternative sensing modality that builds on the same protocol, this time adapted to determining thermal shifts rather than electric noise. To this end, we selectively change the phase of spin $S$ by $\pi/2$ prior to zero-quantum evolution so as to create a time-dependent state

$$\rho_{\text{eff}}(\tilde{\tau}) = \frac{1}{2}(\tilde{S}_- S'_+ + \tilde{S}_+ S'_-)\cos(\delta\omega\tilde{\tau}) - \frac{1}{2i}(\tilde{S}_- S'_+ - \tilde{S}_+ S'_-)\sin(\delta\omega\tilde{\tau}). \quad (4)$$

Note that both terms in Eq. (4) are robust to global magnetic noise, thus making the temporal evolution resilient against this class of fluctuations. After zero- to single-quantum conversion, a temperature change can be calculated as $\delta\mathcal{T} = \delta\omega(d\Delta/d\mathcal{T})^{-1}$, where $\delta\omega$ is the observed frequency shift (in turn, extracted from the signal slope at early times), and $d\Delta/d\mathcal{T}$ is the thermal change in the NV crystal field at room temperature[29].

## 4. Discussion

In summary, we showed that hetero-spin complexes can harbor long-lived coherences arising from transitions at level anti-crossings, or zero-quantum coherences emerging from energy levels associated to magnetic-field-insensitive frequencies. While the former tend to be fragile against field detuning, the latter are long-lived in the presence of strong magnetic noise provided the fields at each spin site remain equal. Under these conditions, zero-quantum coherences have a lifetime insensitive to the magnetic noise amplitude or the dyad's dipolar coupling strength, even though decoherence during longer preparation and readout intervals does lead to a fractional loss of signal contrast. To mitigate this problem, one could resort to focused implantation of molecular nitrogen or related techniques, already explored as a strategy to producing multi-NV clusters[36,37].

Although we assumed herein an isolated electron spin dyad, a color center such as the NV would typically interact with not one but several paramagnetic impurities, themselves interconnected to form an electronic spin bath. Flip-flops between spins in the bath would lead to diffusion and ultimately to a loss of coherence in the entangled spin dyad. These dynamics, however, rest on energy matching between proximal spins, implying the process can be countered by introducing sufficiently large spectral shifts, attained, e.g., through magnetic field gradients[38-40]. Alternatively, one can resort to selective radio-frequency excitation and polarization transfer[41] to initialize the nuclear hosts of surrounding spin-1/2 centers into different hyperfine states so that only one is resonant at the applied magnetic field.

Other than the spin dyad formed by an NV and a neighboring spin-1/2 center, similar dynamics are to be expected for hetero-spin pairs where the response of an individual spin to a global field change counters the other. Examples include pairs comprising a spin-1/2 center — which tend to be ubiquitous — and other optically addressable spin qubits in diamond, silicon carbide, or silicon, to mention only a few material hosts attracting present interest.

## Acknowledgements

D.P. and C.A.M. acknowledge support from the National Science Foundation through grants NSF-1903839 and NSF-2203904. P.R.Z. acknowledges support from SeCyT-UNC through grant 33620180100154CB and CONICET through grant PIP 11220200102451CO. All authors also acknowledge access to the facilities and research infrastructure of the NSF CREST IDEALS, grant number NSF-2112550.

## Conflict of Interest

The authors declare no conflict of interest.

## Data Availability Statement

The data that support the findings of this study are available from the corresponding author upon reasonable request.

# Supplementary Material for

# "Quantum sensing via magnetic-noise-protected states in an electronic spin dyad"

Carlos A. Meriles[1,2,†], Pablo R. Zangara[3,4], and Daniela Pagliero[1]

[1]Department. of Physics, CUNY-City College of New York, New York, NY 10031, USA.
[2]CUNY-Graduate Center, New York, NY 10016, USA.
[3]Universidad Nacional de Córdoba, Facultad de Matemática, Astronomía, Física y Computación, Córdoba, Argentina.
[4] CONICET, Instituto de Física Enrique Gaviola (IFEG), Córdoba, Argentina.

[†]Corresponding author. E-mail: cmeriles@ccny.cuny.edu.

**I-Spin Hamiltonian**

We consider an electronic spin dyad comprising a spin $S = 1$ with a crystal field $\Delta$ and a neighboring paramagnetic impurity $S' = 1/2$. Assuming the magnetic field $B$ is parallel to the crystal field axis, we write the system Hamiltonian as

$$H = \Delta S_z^2 + |\gamma_e|B S_z + |\gamma_e|B S'_z + H_d, \qquad (S1)$$

where $H_d$ represents the dipolar interaction given by

$$H_d = 2\pi \mathcal{J} \left\{ (1 - 3\cos^2\theta)\left(S_z S'_z - \frac{1}{4}(S_+ S'_- + S_- S'_+)\right) \right.$$
$$\left. - \frac{3}{4}\sin 2\theta \left((S_+ + S_-)S'_z + S_z(S'_+ + S'_-)\right) - \frac{3}{4}\sin^2\theta \,(S_+ S'_+ + S_- S'_-)\right\}. \qquad (S2)$$

In Eq. (S2), the coupling amplitude is given by $2\pi \mathcal{J} = \frac{\mu_0 \gamma_e^2 \hbar^2}{4\pi r^3}$, $\mu_0$ is the vacuum permeability, $\hbar$ is the reduced Planck constant, $\gamma_e$ is the electronic gyromagnetic ratio, $\Delta$ is the crystal field (expressed in rad s$^{-1}$), $r$ denotes the inter-spin separation, and $\theta$ is the angle formed by the inter-spin vector and the magnetic field; we also use the standard notation for the ladder operators $S_\pm = S_x \pm iS_y$ and similarly for spin $S'$.

Provided the microwave (MW) excitation of spin $S$ is limited to address the transition between the $|m_S = 0\rangle$ and $|m_S = -1\rangle$ states, we restrict our description to the manifold formed by states $|1\rangle = |0, +1/2\rangle$, $|2\rangle = |-1, +1/2\rangle$, $|3\rangle = |0, -1/2\rangle$, and $|4\rangle = |-1, -1/2\rangle$, and describe spin $S$ via a fictitious spin-1/2 operator $\tilde{S}$. In this representation, we describe the NV via the virtual spin $\tilde{S} = 1/2$ and recast $S_z$ as

$$S_z \to \tilde{S}_z - \mathbb{I}/2 \qquad (S3)$$

The Hamiltonian then takes the simpler form

$$H = (|\gamma_e|B - \Delta)\tilde{S}_z + (|\gamma_e|B - \pi \mathcal{J}_\parallel)S'_z + 2\pi \mathcal{J}_\parallel \tilde{S}_z S'_z + 2\pi \mathcal{J}_\perp \sqrt{2}(\tilde{S}_+ S'_+ + \tilde{S}_- S'_-), \qquad (S4)$$

where we ignore contributions proportional to the identity operator $\mathbb{I}$. The last two terms capture the secular contributions of the dipolar interaction with the notation $\mathcal{J}_\parallel = \mathcal{J}(1 - 3\cos^2\theta)$ and $\mathcal{J}_\perp = -\frac{3}{4}\mathcal{J}\sin^2\theta$. Equation (S4) yields a level anti-crossing[*] at $|\gamma_e|B_m = (\Delta + \pi \mathcal{J}_\parallel)/2$ where states $|1\rangle$ and $|4\rangle$ hybridize to yield the eigenstates $|\pm\rangle = 1/\sqrt{2}\,(|1\rangle \pm |4\rangle)$. Far enough from $B_m$, the last term becomes non-secular and can be ignored, thus leading back to the expression in Eq. (2) of the main text[1].

---

[*] Note that observation of the level anti-crossing requires a spin pair whose coupling is stronger than the relaxation rate of either spin in the dyad, a condition most likely met in isolated pairs. By contrast, the energy gap will be washed out in a system where spin $S'$ is part of a bath as the coupling with other (like) spins in the ensemble (statistically equivalent to that with spin $S$) introduces a line broadening comparable to the level splitting.

Lastly, we note that in the presence of MW, Eq. (S4) must be supplemented with a term of the form $H_{MW} = \sqrt{2}|\gamma_e|B_1\tilde{S}_x \cos\omega t + |\gamma_e|B_1'S_x' \cos\omega' t$, where $B_1$ and $B_1'$ denote the MW field amplitudes at frequencies $\omega$ and $\omega'$, respectively resonant with spins $S$ and $S'$. At the level anti-crossing, $\omega \approx \omega'$ with the practical consequence that the MW manipulation of spins $S$ and $S'$ now relies on a common field of amplitude $B_1 = B_1'$. Correspondingly, rotations of either spin species cannot be controlled independently, and the effectiveness of pulse sequences is negatively impacted (for example, in the Hahn-echo protocol of Fig. 2 in the main text, a rotation by an angle $\theta$ of spin $S$ amounts to a rotation $\theta/\sqrt{2}$ for spin $S'$). The latter, however, represents only a minor complication in the sense that although the overall signal contrast necessarily shrinks, imperfect spin rotations have no impact on the duration of the spin coherences, hence making the field-dependent results in Fig. 2 of the main text still valid. Further, the differing MW amplitudes at $B_m$ can simply be seen as the result of "MW field heterogeneity", and thus its effect can be efficiently mitigated, e.g., by resorting to composite pulses.

For clarity, we ignore in Eq. (S1) any hyperfine couplings of either paramagnetic center with their nuclear spin host This simplification is valid so long as the nuclear spin lifetime is longer than the protocol duration, a condition met in most solid-state spin qubits. The latter also applies to systems exhibiting a dynamic Jahn-Teller distortion provided the process is sufficiently slow; specifically, this is the case of the P1 center, a system with $C_{3v}$ symmetry whose room-temperature reorientation — and concomitant change of the hyperfine coupling — takes place on a scale of several seconds[2,3].

## II-Polarization transfer

Assuming optical spin initialization of spin $S$ into the $|m_S = 0\rangle$ state, we write the system density matrix as

$$\rho(0) = |0\rangle\langle 0| = \frac{\mathbb{I}}{4} + \frac{\tilde{S}_z}{2}. \tag{S5}$$

Following the polarization transfer protocol in Fig. 3a, the state at a time $2\tau_{ZQ}^{(-)}$ (i.e., immediately before the $(\pi/2)_y$ pulse) takes the form

$$\rho\left(2\tau_{ZQ}^{(-)}\right) = \frac{\mathbb{I}}{4} + \tilde{S}_x S_z', \tag{S6}$$

where we used the relation $\exp(-i2\pi J_\parallel \tilde{S}_z S_z' t)\tilde{S}_y \exp(i2\pi J_\parallel \tilde{S}_z S_z' t) = \tilde{S}_y \cos(\pi J_\parallel t) - 2\tilde{S}_x S_z' \sin(\pi J_\parallel t)$ and the condition $\tau_{ZQ} = (4J_\parallel)^{-1}$. Upon application of a $(\pi/2)_y$ pulse, the density matrix becomes

$$\rho\left(2\tau_{ZQ}^{(+)}\right) = \frac{\mathbb{I}}{4} - \tilde{S}_z S_x', \tag{S7}$$

and following evolution during the second half of the protocol, we obtain

$$\rho\left(4\tau_{ZQ}^{(+)}\right) = \frac{\mathbb{I}}{4} - \frac{S_z'}{2}. \tag{S8}$$

Re-pumping spin $S$ into $|m_S = 0\rangle$, the density matrix takes the final form

$$\rho_{\text{Init}} = \left(\frac{\mathbb{I}}{2} + \tilde{S}_z\right)\left(\frac{\mathbb{I}}{2} - S_z'\right) = \frac{1}{4}\left(\mathbb{I} + 2(\tilde{S}_z - S_z') - 4\tilde{S}_z S_z'\right) = |0, -1/2\rangle\langle -1/2, 0|. \tag{S9}$$

## III-Coherence order conversion

For a pulse sequence of the form $(\pi/2)_x \rightarrow (\pi)_x \rightarrow (\pi/2)_x$ (lower panel in Fig. 3a), the evolution operator can be expressed as

$$U_{\text{COC}} = \exp\left(-i\frac{\pi}{2}(\tilde{S}_x + S_x')\right)\exp(-H\tau_{ZQ})\exp\left(-i\pi(\tilde{S}_x + S_x')\right)\exp(-H\tau_{ZQ})\exp\left(-i\frac{\pi}{2}(\tilde{S}_x + S_x')\right)$$

$$= \exp\left(+i\frac{\pi}{2}(\tilde{S}_x + S_x')\right)\exp(-i2\pi J_\parallel \tilde{S}_z S_z' 2\tau_{ZQ})\exp\left(-i\frac{\pi}{2}(\tilde{S}_x + S_x')\right)$$

$$= \exp(-i2\pi\mathcal{J}_\parallel \tilde{S}_y S'_y 2\tau_{ZQ}), \qquad . \tag{S10}$$

where we have assumed $B \neq B_{\text{m}}$. Therefore, if Eq. (S9) describes the system state after initialization, the density matrix after coherence order conversion (COC) can be cast as

$$\rho_{ZQ} = \frac{1}{4}\left(\mathbb{I} + 2U_{\text{COC}}(\tilde{S}_z - S'_z)U^\dagger_{\text{COC}} - 4U_{\text{COC}}\tilde{S}_z S'_z U^\dagger_{\text{COC}}\right)$$

$$= \frac{1}{4}\left(\mathbb{I} + 4(\tilde{S}_x S'_y - \tilde{S}_y S'_x) - 4\tilde{S}_z S'_z\right).$$

$$= \frac{1}{4}\left(\mathbb{I} - 2i(\tilde{S}_- S'_+ - \tilde{S}_+ S'_-) - 4\tilde{S}_z S'_z\right). \tag{S11}$$

We note that the term $\tilde{S}_z S'_z$ in the expression for $\rho_{\text{Init}}$ is insensitive to $U_{\text{COC}}$ hence allowing us to define an effective density matrix $\rho_{\text{eff}}$ that only takes into account the contribution deriving from the $(\tilde{S}_z - S'_z)$ term; this is the approach we follow in the main text.

**IV-Evolution of zero-quantum coherences**

In the simplest scenario, the system evolves freely without any external excitation (we also assume $B \neq B_{\text{m}}$). The first and last terms in Eq. (S11) commute with $H$ and hence undergo no evolution; we therefore write

$$\rho_{ZQ,\text{eff}}(\tau) = \rho_{ZQ} - \frac{1}{4}\left(\mathbb{I} - 4\tilde{S}_z S'_z\right) = \frac{1}{2i}\exp(-iH\tau)\left(\tilde{S}_- S'_+ - \tilde{S}_+ S'_-\right)\exp(iH\tau)$$

$$= \frac{1}{2i}\exp(-i2\pi\mathcal{J}_\parallel \tilde{S}_z S'_z \tau)\left(\tilde{S}_- S'_+ - \tilde{S}_+ S'_-\right)\exp(i2\pi\mathcal{J}_\parallel \tilde{S}_z S'_z \tau), \tag{S12}$$

where we can ignore the terms in $H$ linear in $\tilde{S}_z$, $S'_z$ after a double rotating frame transformation resonant with spins $\tilde{S}$ and $S'$. After a bit of algebra, one can prove that

$$\left[(\tilde{S}_- S'_+ - \tilde{S}_+ S'_-), \tilde{S}_z S'_z\right] = 2i\left[(\tilde{S}_x S'_y - \tilde{S}_y S'_x), \tilde{S}_z S'_z\right] = 0 \tag{S13}$$

implying that

$$\rho_{ZQ,\text{eff}}(\tau) = \frac{1}{2i}(\tilde{S}_- S'_+ - \tilde{S}_+ S'_-) = (\tilde{S}_x S'_y - \tilde{S}_y S'_x), \tag{S14}$$

i.e., independent of time. To assess the impact of noise, we consider a contribution to the Hamiltonian of the form $H_n = |\gamma_e|(\beta\tilde{S}_z + \beta' S'_z)$ with $\beta$ and $\beta'$ constant, and make use of standard spin transformation rules to re-calculate $\rho_{ZQ,\text{eff}}$ after an evolution interval $\tau$; we find

$$(\tilde{S}_x S'_y - \tilde{S}_y S'_x) \to (\tilde{S}_x S'_y - \tilde{S}_y S'_x)\cos(|\gamma_e|(\beta - \beta')\tau) + (\tilde{S}_x S'_x + \tilde{S}_y S'_y)\sin(|\gamma_e|(\beta - \beta')\tau) \tag{S15}$$

From the expression above, it is clear that the system dephases in the presence of imbalance between the noise amplitudes at either spin site (the case for $\beta_S(t)$ and $\beta_{S'}(t)$ in the main text). Partial compensation can be attained, e.g., by intercalating a $\pi$-pulse at the midpoint of the zero-quantum evolution interval (or, more generally, by a train of inversion pulses). This can be seen from the fact that a (global) $(\pi)_x$-pulse inverts the sign of $(\tilde{S}_x S'_y - \tilde{S}_y S'_x)$ but leaves $(\tilde{S}_x S'_x + \tilde{S}_y S'_y)$ unchanged. Further, after a similar derivation we find

$$(\tilde{S}_x S'_x + \tilde{S}_y S'_y) \to (\tilde{S}_x S'_x + \tilde{S}_y S'_y)\cos(|\gamma_e|(\beta - \beta')\tau) - (\tilde{S}_x S'_y - \tilde{S}_y S'_x)\sin(|\gamma_e|(\beta - \beta')\tau) \tag{S16}$$

Therefore, evolution under a protocol of the form $\tau - (\pi)_x - \tau$ yields

$$\rho_{ZQ,\text{eff}}(\tau,\tau') = -(\tilde{S}_x S'_y - \tilde{S}_y S'_x)\cos(|\gamma_e|(\beta - \beta')(\tau - \tau')) + (\tilde{S}_x S'_x + \tilde{S}_y S'_y)\sin(|\gamma_e|(\beta - \beta')(\tau - \tau')) \tag{S17}$$

hence leading to an echo at $\tau = \tau'$. In this sense, the present approach is fully compatible with dynamical decoupling protocols, which can be integrated into the protocol for additional noise protection.

For completeness, we mention that electric fields $\vec{\mathcal{E}}$ selectively affect spin $S$ through contributions to the Hamiltonian of the form

$$H_\varepsilon = \delta_\| \mathcal{E}_z \left(S_z^2 - \frac{2}{3}\right) - \delta_\perp \left(\mathcal{E}_x(S_x S_y + S_y S_x) + \mathcal{E}_y(S_x^2 - S_y^2)\right) \rightarrow$$
$$-d_\| \mathcal{E}_z \tilde{S}_z - 2\delta_\perp \left(\mathcal{E}_x(\tilde{S}_x \tilde{S}_y + \tilde{S}_y \tilde{S}_x) + \mathcal{E}_y(\tilde{S}_x^2 - \tilde{S}_y^2)\right), \quad (S18)$$

where the last expression holds in the reduced representation used herein with the correspondence $S_{x,y} \rightarrow \sqrt{2}\, \tilde{S}_{x,y}$ (see Section I). Since spin $S'$ is insensitive to electric fields, no phase compensation takes place and the dyad behaves as an electrometer, as stated in the main text.

Finally, thermal sensing in Fig. 4 starts with a π/2-phase shift on spin $\tilde{S}$, hence leading to the transformation

$$(\tilde{S}_x S'_y - \tilde{S}_y S'_x) \rightarrow (\tilde{S}_y S'_y + \tilde{S}_x S'_x) \rightarrow (\tilde{S}_y S'_y + \tilde{S}_x S'_x)\cos(\delta\omega\tilde{\tau}) - (\tilde{S}_x S'_y - \tilde{S}_y S'_x)\sin(\delta\omega\tilde{\tau}) \quad (S19)$$

where we obtain the last expression after evolution for a time $\tilde{\tau}$ under the Hamiltonian $H(\mathcal{T}) = \delta\omega \tilde{S}_z + 2\pi \mathcal{J}_\| \tilde{S}_z S'_z$, with $\delta\omega = \left(\frac{d\Delta}{d\mathcal{T}}\right)\delta\mathcal{T}$. Note that the first term in the sum, $(\tilde{S}_y S'_y + \tilde{S}_x S'_x) = \frac{1}{2}(\tilde{S}_+ S'_- + \tilde{S}_- S'_+)$, remains unchanged (and thus undetectable) upon application of a zero- to single-quantum conversion whereas the second term transforms as $(\tilde{S}_x S'_y - \tilde{S}_y S'_x) \rightarrow \frac{1}{2}(\tilde{S}_z - S'_z)$. Therefore, observation of spin $\tilde{S}$ yields a net signal

$$\Sigma = -\frac{1}{2}\text{Tr}\{\tilde{S}_z^2\}\sin(\delta\omega\tilde{\tau}) \approx -\frac{1}{4}\delta\omega\tilde{\tau}, \quad (S20)$$

where the last expression is valid at sufficiently short evolution times.

**V-Spin dynamics simulations**

Unitary spin dynamics is evaluated by exact diagonalization of the Hamiltonian in Eq. (S4). In order to include the effects of the magnetic 'fluctuators', we add a time-dependent noise amplitude to the external field, $B \rightarrow B + \beta(t)$. Here, the random variable $\beta(t)$ changes over time at a fixed rate $r$. Every time it changes, it takes a value from a uniform distribution with zero mean and width $\beta_{\text{rms}}$. In practice, the quantum dynamics of the spin system is computed by small time-steps $dt$, so the noise-induced field shifts occur with a probability $p = r \times dt$. Averaging over many trajectories, i.e. the full evolution of a complete pulse sequence, yields the final density matrix.

To gauge the impact of local differences between the environments at either spin site, we model the noise fields as the sum of a global and a local contribution independent from each other, namely, $\beta(t) = \beta_g(t) + \beta_l(t)$ and $\beta'(t) = \beta_g(t) + \beta'_l(t)$. In our simulations, we choose $\beta_l(t)$ and $\beta'_l(t)$ so that $\beta_{\text{rms}} = \beta'_{\text{rms}} = \text{constant}$. Under these conditions $\langle \beta_l^2 \rangle = \langle \beta'^2_l \rangle$, and $\xi \rightarrow 0$ as $\langle \beta_l^2 \rangle \rightarrow 0$ (correspondingly, $\xi \rightarrow 1$ as $\langle \beta_g^2 \rangle \rightarrow 0$). In our simulations, all pulses are considered instantaneous and perfect transformations, so there is no magnetic noise acting during them.